\documentclass[twocolumn]{aastex631}

\usepackage{natbib}
\usepackage{multirow}
\usepackage{booktabs}
\usepackage{epsfig}
\usepackage{ulem}
\usepackage{rotating}
\usepackage{graphicx}
\usepackage{url}
\usepackage{amsmath}
\usepackage{color}
\usepackage{savesym}
\savesymbol{tablenum}
\usepackage{siunitx}
\usepackage[normalem]{ulem}
\usepackage{appendix}
\usepackage{enumerate}
\usepackage{gensymb}
\usepackage{fontawesome}
\usepackage[mathlines]{lineno}

\usepackage{longtable}
\usepackage{afterpage}
\usepackage{threeparttable}
\usepackage{tabularx}
\usepackage[bottom]{footmisc}

\DeclareSIUnit\parsec{pc}
\DeclareSIUnit\h{\textit{h}}

\defcitealias{BroutScolnic21}{BS21}
\defcitealias{Popovic2025}{Dovekie}
\defcitealias{Brout_2022}{Fragilistic}

{{\bm{\hat{\textnormal{\bfseries\i}}}}}

\DeclareRobustCommand{\uvec}[1]{{%
  \ifcsname uvec#1\endcsname
     \csname uvec#1\endcsname
   \else
    \bm{\hat{\mathbf{#1}}}%
   \fi
}}
\newcommand{\git}[2]{\href{https://github.com/#1/#2}{\texttt{#2}~\faGithub}\footnote{\url{https://github.com/#1/#2}}}

\newcommand{\gitalias}[3]{
{\href{https://github.com/#1/#2}{\texttt{#3}~\faGithub}\footnote{\url{https://github.com/#1/#2}}}
}

\begin{document}
\title{The Dark Energy Bedrock All-Sky Supernova Program: Cross Calibration, Simulations, and Cosmology Forecasts}

    \author[0000-0002-5389-7961]{Maria Acevedo}
    \affiliation{Department of Physics, Duke University, Durham, NC 27708, USA}
    \author[0000-0001-5399-0114]{Nora F. Sherman}
    \affiliation{Institute for Astrophysical Research, Boston University, 725 Commonwealth Avenue, Boston, MA 02215, USA}
    \author[0000-0001-5201-8374]{Dillon Brout}
    \affiliation{Departments of Astronomy and Physics, Boston University, 725 Commonwealth Avenue, Boston, MA 02215, USA}
    \author[0000-0002-7234-844X]{Bastien Carreres}
    \affiliation{Department of Physics, Duke University, Durham, NC 27708, USA}
    \author[0000-0002-4934-5849]{Daniel Scolnic}
    \affiliation{Department of Physics, Duke University, Durham, NC 27708, USA}
    \author[0000-0002-8012-6978]{Brodie Popovic}
    \affiliation{Universite Claude Bernard Lyon 1, CNRS, IP2I Lyon IN2P3, IMR 5822, F-69622 Villeurbanne, France}
    \author[0000-0003-1997-3649]{Patrick Armstrong}
    \affiliation{The Research School of Astronomy and Astrophysics, The Australian National University, Canberra, ACT 2611, Australia}
    \author[0009-0006-5649-5067]{Dingyuan Cao}
    \affiliation{The Research School of Astronomy and Astrophysics, The Australian National University, Canberra, ACT 2611, Australia}
    \author[0000-0002-4934-5849]{Rebecca C. Chen}
    \affiliation{Department of Physics, Duke University, Durham, NC 27708, USA}
    \author[0000-0001-8251-933X]{Alex Drlica-Wagner}
    \affiliation{Fermi National Accelerator Laboratory, P.O.\ Box 500, Batavia, IL 60510, USA}
    \affiliation{Kavli Institute for Cosmological Physics, University of Chicago, Chicago, IL 60637, USA}
    \affiliation{Department of Astronomy and Astrophysics, University of Chicago, Chicago, IL 60637, USA}
    \affiliation{NSF-Simons AI Institute for the Sky (SkAI),172 E. Chestnut St., Chicago, IL 60611, USA}
    \author[0000-0001-6957-1627]{Peter S. Ferguson}
    \affiliation{DiRAC Institute, Department of Astronomy, University of Washington, 3910 15th Ave NE, Seattle, WA, 98195, USA}
    \author[0000-0003-1731-0497]{Christopher Lidman}
    \affiliation{The Research School of Astronomy and Astrophysics, The Australian National University, Canberra, ACT 2611, Australia}
    \author[0009-0006-4963-3206]{Bailey Martin}
    \affiliation{The Research School of Astronomy and Astrophysics, The Australian National University, Canberra, ACT 2611, Australia}
    \author[0000-0001-8596-4746]{Erik R.~Peterson}
    \affiliation{Department of Physics, Duke University, Durham, NC 27708, USA}
    \author[0000-0002-6124-1196]{Adam G.~Riess}
    \affiliation{Space Telescope Science Institute, Baltimore, MD 21218, USA}
    \affiliation{Department of Physics and Astronomy, Johns Hopkins University, Baltimore, MD 21218, USA}

\begin{abstract}
Type Ia supernovae (SNe~Ia) have been essential for probing the nature of dark energy; however, most SN analyses rely on the same low-redshift sample, which may lead to shared systematics. In a companion paper \citep{nora}, we introduce the Dark Energy Bedrock All-Sky Supernova (DEBASS) program, which has already collected more than 500 low-redshift SNe~Ia on the Dark Energy Camera (DECam), and present an initial release of 77 SNe~Ia within the Dark Energy Survey (DES) footprint observed between 2021 and 2024. Here, we examine the systematics, including photometric calibration and selection effects. We find agreement at the 10 millimagnitude level among the tertiary standard stars of DEBASS, DES, and Pan-STARRS1. Our simulations reproduce the observed distributions of DEBASS SN light-curve properties, and we measure a bias-corrected Hubble residual scatter of $0.08$~mag, which, while small, is found in $10\%$ of our simulations. We compare the DEBASS SN distances to the Foundation sample and find consistency with a median residual offset of $0.016 \pm 0.019$~mag. Selection effects have negligible impacts on distances, but a different photometric calibration solution shifts the median residual $-0.015 \pm 0.019$~mag, highlighting calibration sensitivity. Using conservative simulations, we forecast that replacing historical low-redshift samples with the full DEBASS sample ($>$400 SNe~Ia) will improve the statistical uncertainties on dark energy parameters $w_0$ and $w_a$ by 30\% and 24\% respectively, enhance the dark energy Figure of Merit by up to 60\%, and enable a measurement of $f\sigma_8$ at the 25\% level.

\end{abstract}

\section{Introduction}
In cosmological studies with Type Ia supernovae (SNe~Ia), systematic uncertainties have long rivaled statistical uncertainties, underscoring the need for precise characterization of systematics to improve cosmological measurements \citep{Betoule14, des2019, Scolnic18, Popovic2025}. This is particularly significant in light of recent analyses of Baryonic Acoustic Oscillations (BAO; \citealp{Giar__2024, desidr2_bao}) and SNe~Ia \citep{Brout22, vincenzi2024, union}, which offer potential evidence for a time-evolving dark energy. Although these independent SN analyses have different high-$z$ SNe and show relatively consistent results, they rely on the same low-redshift samples. The Dark Energy Bedrock All-Sky Supernova (DEBASS) program aims to provide a uniform low-$z$ replacement and mitigate systematics from heterogeneous calibrations. In a companion paper, we release the first DEBASS data (DR0.5), showcasing the program’s use of DECam's well-established observational and image-reduction framework \citep{nora}. Here, we make the first attempt to quantify the uncertainties within our new dataset that differ from those used on the same telescope for the Dark Energy Survey Supernova Program (DES; \citealp{vincenzi2024}), and we forecast cosmological constraints for a full DEBASS+DES analysis covering $0.001 \lesssim z \lesssim 1.2$ on the same telescope.

One of the current challenges in assembling a low-$z$ SN~Ia sample is the need to use multiple surveys with varying instrumentation, calibration standards, and selection criteria. For example, significant efforts were made for the Pantheon+ sample to address discrepancies such as zeropoint offsets and wavelength-dependent filter shifts, requiring extensive cross-calibration of the data \citep{Scolnic15, Brout_2022}. These calibration inconsistencies can propagate into biases in cosmological parameters \citep{Popovic2025}. Additionally, the selection effects of the older literature samples are especially difficult to model, further complicating analyses \citep{Scolnic18}.

The Dark Energy Survey addressed many of these challenges by conducting a six-year (2013--2019) optical imaging program with the Dark Energy Camera (DECam) on the 4-m Blanco telescope at Cerro Tololo Inter-American Observatory. DES mapped $\sim$5000~deg$^2$ of the southern sky and monitored ten deep fields to build the DES-SN5YR sample of $\sim$1800 SNeIa from $0.1<z<1.2$ \citep{sanchez2024darkenergysurveysupernova}. A key advantage of DEBASS is its ability to leverage the DECam calibration infrastructure. The Forward Global Calibration Method (FGCM) models time- and position-dependent variations in atmospheric transmission, instrumental throughput, and CCD response, placing all DES imaging on a uniform photometric scale \citep{Burke_2017, rykoff2023darkenergysurveysixyear}. Additional cross-calibrations tie the DES system to spectrophotometric standards and Pan-STARRS1 \citep{Scolnic15, Brout_2022}. The impact of filter-dependent and atmospheric corrections is discussed in \citet{Lasker_2019}. 

To increase the usable sky area for DEBASS SN follow-up beyond that of the $\sim$5000~deg$^2$ DES wide-field survey, we combine DECam cross-calibration efforts using data from the DECam Local Volume Exploration (DELVE; \citealp[]{DW_2022}) survey and other DECam-based programs. 
By anchoring our calibration to this framework—and validating it with independent checks—DEBASS minimizes cross-program zeropoint uncertainties and enables a more homogeneous low-$z$ SN~Ia sample. When combined with the complementary high-redshift, photometrically-classified SN~Ia sample from DES-SN5YR \citep{vincenzi2024}, we will have one of the largest single-instrument SN~Ia cosmology samples spanning $0.001 \lesssim z \lesssim 1.2$. We note that a similar single-instrument strategy was implemented by the Foundation Supernova Survey, which used the Pan-STARRS1 (PS1) telescope and photometric system, demonstrating the benefits of coherent calibration and reduced cross-program systematics for precision cosmology \citep{Foley18}. A joint cosmological analysis of the Foundation sample with the PS1-MD sample \citep{Scolnic18} was done in \citet{Jones19}.

\begin{table*}[!ht] 
    \centering
    \caption{Calibration offsets (in magnitudes) between observed and synthetic stellar magnitudes in each band. For DEBASS, offsets represent the vertical shift of the linear best fit to the stellar comparison sample. For \citetalias{Brout_2022} and \citetalias{Popovic2025}, the offsets are the difference between the calibration offsets for DES-SN5YR and PS1.}
    \vspace{-0.2em}
    \begin{tabularx}{\textwidth}{l@{\extracolsep{\fill}}|ccccc}
    & $g$     & $r$     & $i$     & $z$ & Description    \\\hline\hline    
    Acevedo et al. (2025)~~~ & $0.006 \pm 0.007$ & $-0.022 \pm 0.007$ & $-0.015 \pm 0.007$ & $-0.025 \pm 0.007$ & This work\\
    \citetalias{Brout_2022} & $0.012 \pm 0.007$  & $-0.014 \pm 0.007$ & $-0.009 \pm 0.007$ & $-0.007 \pm 0.007$ & Tertiary stars\\
    \citetalias{Popovic2025} & $0.003 \pm 0.004$ & $-0.005 \pm 0.004$ & $0.003 \pm 0.004$ & $-0.017 \pm 0.005$ & DA white dwarfs\\\hline
    \end{tabularx}
    \label{tab:cal_offsets}
\end{table*}

Recent SN analyses have dedicated significant effort to developing accurate simulations, leveraging them to refine cosmological measurements by accounting for expected biases \citep{Kessler_2017, taylor2023}, test different analyses \citep{qu2024,Armstrong_2023,Camilleri_2024,carreresTypeIaSupernova2025}, develop physically motivated model of intrinsic variations \citep{BroutScolnic21,Popovic_2021,popovic2022,wiseman_2022}, and more. A key component of these simulations is a well-characterized selection function, which is essential for interpreting the observed SN population and correcting for selection biases. DEBASS is well-positioned to meet this need by targeting transients discovered by discovery surveys such as Asteroid Terrestrial-impact Last Alert System (ATLAS; \citealp[]{ATLAS}), All-Sky Automated Survey for Supernovae (ASAS-SN; \citealp[]{asassn}), and the Zwicky Transient Facility (ZTF; \citealp[]{Bellm19}). Follow-up targets are selected based on redshift ($0.01 < z < 0.08$, or in a potential Distance Ladder host), low Galactic extinction ($E(B-V)<0.25$~mag), and visibility with DECam for at least 30 days, enabling the capture of well-sampled and high signal-to-noise ratio light curves \citep{nora}. The DEBASS program was awarded an average of four total nights per semester, spread across $\sim50$ epochs each semester with a typical cadence of 3 days. DEBASS uses an adaptive cadence, increasing observation frequency near maximum light (with 88\% of DR0.5 having pre-peak coverage and 74\% with $\geq5$ days pre-peak data). To further address modeling uncertainties, we present the first set of simulations designed to quantify and reduce systematics in our analysis.

Although the initial DEBASS data release is much smaller than the eventual program sample of over 500 SNe~Ia, it offers valuable opportunities for validation and consistency checks with previous low-$z$ datasets. In particular, we compare our results to the Foundation Supernova Survey \citep{Foley18}, which currently serves as the dominant low-$z$ anchor for nearly all cosmological analyses \citep{Brout22}. Comparing DEBASS to Foundation allows us to assess how systematic differences in calibration, filter systems, and selection strategies may influence cosmological inferences. Whether DEBASS agrees with or diverges from this dataset on the Hubble diagram has significant implications for the robustness of SN~Ia cosmology with this sample and the reliability of our low-$z$ constraints. These early findings not only establish a benchmark for the program but also highlight the potential for DEBASS to refine cosmological measurements with SNe~Ia as the dataset grows.

The paper is organized as follows. In Section~\ref{sec:Calib}, we describe the photometric calibration of the DEBASS sample, including both internal consistency checks and external cross-program comparisons. In Section~\ref{sec:Sims}, we present our simulation framework: we detail the light-curve fitting process, construction of simulations, modeling of the selection function, and comparisons between simulated and observed data, culminating in the derivation of bias corrections. Section~\ref{sec:results} focuses on key results, including systematic effects on distance estimates, Hubble residuals, and the observed Hubble scatter. Finally, in Section~\ref{sec:forecast}, we provide cosmological forecasts based on the full DEBASS sample, including constraints on the growth rate of structure and the time evolution of dark energy. We conclude with a discussion and summary in Section~\ref{sec:Conclusions}.

\section{Calibration}\label{sec:Calib}
\subsection{Internal Calibration}\label{sec:intcalib}
Our dataset's photometric calibration begins with single-epoch processed image files. We extract a catalog of detected objects from these images, identifying stars suitable for calibration. Our calibration process then proceeds by cross-matching these objects with existing reference catalogs, leveraging the well-characterized photometry from DELVE DR2 as our primary reference \citep{DW_2022}.

Although the DEBASS DR0.5 release focuses on supernovae discovered within the DES footprint, the full DEBASS program spans a much larger area across the southern sky. To ensure consistency across this broader program, we adopt DELVE DR2 as our primary photometric reference for all fields. For objects found in the DELVE DR2 catalog, we compare their magnitudes to those measured in our images. The difference between the DELVE magnitudes and our magnitudes provides the zeropoint (ZP) for the image, allowing us to account for variations in atmospheric transparency, instrumental throughput, and other observational factors. Since DELVE adopts the DES calibration within the DES footprint, the two surveys can be treated as interchangeable in this region. Outside of the DES footprint, DELVE uses the ATLAS Refcat2 catalog to derive zeropoints \citep{DW_2022, tonry18}. We confirm this by comparing the nightly photometry of our calibration stars with zeropoints calibrated by DELVE, to the DES Y6 standard star catalog \citep{rykoff2023darkenergysurveysixyear}. We find mean residuals of $0.0016 \pm 0.0020$~mag in $g$, $-0.0004 \pm 0.0009$~mag in $r$, $-0.0016 \pm 0.0023$~mag in $i$, and $0.0003 \pm 0.0012$~mag in $z$. These offsets are all below 2 millimagnitudes. Although these shifts are negligible for our analysis, we nonetheless include them as a systematic uncertainty in Section~\ref{sec:results}.

\subsection{External Photometric Calibration}\label{sec:extcalib}

To ensure consistent comparisons between our sample and other existing SN~Ia datasets, we compare the stellar magnitudes observed by DELVE for DECam to those of  PS1 in the region where the two programs overlap ($-30^\circ <$~RA~$< 30^\circ$). We obtain PS1 aperture magnitudes from the PS1 DR2 \citep{ps1dr2} and DELVE magnitudes from DELVE DR2 \citep{DW_2022}, matching sources based on RA and DEC with a one-arcsecond tolerance. Each star is corrected for possible Milky Way extinction using IRSA6,\footnote{\url{https://irsa.ipac.caktech.edu/}.} as implemented in \citet{Popovic2025}, based on the maps from \citet{Schlafly_2011}. The correction is applied at the effective wavelength of each filter. Following the approach of \citet{Brout_2022} and \citet{Popovic2025}, we apply magnitude cuts to ensure a linear photometric response in the PS1 system: $g > 14.8$~mag, $r > 14.9$~mag, $i > 15.1$~mag, and $z > 14.6$~mag. To mitigate Malmquist bias, we further restrict the PS1 $g$-band to be brighter than 19 magnitudes. Additionally, we limit the stellar color range to $0.25~\mathrm{mag} < g - i < 1.0~\mathrm{mag}$. Table~\ref{tab:cal_offsets} details the magnitude offset between the linear best fit to our data in each band relative to the expectation from synthetic spectra (\citealp{ngsl}), which are of comparable size to those reported in other cross-program calibration efforts such as \citet{Brout_2022} (hereafter \citetalias{Brout_2022}) and \citet{Popovic2025} (hereafter \citetalias{Popovic2025}). \citetalias{Brout_2022} provided a unified cross-calibration of 25 photometric systems using Pan-STARRS stellar photometry, solving for filter offsets and propagating their uncertainties into light-curve retraining and cosmological fits. \git{bap37}{Dovekie} improves upon Fragilistic by introducing an open-source framework that incorporates DA white dwarf standards and directly modifies filter transmission curves based on stellar observations, leading to improved calibration precision \citep{Popovic2025}. While the specific offsets are not identical across studies \textemdash likely reflecting differences in stellar samples (which have been shown to vary by 5~mmag across the footprint), assumed effective filter wavelength, and calibration methodology \textemdash the band‑to‑band agreement is generally good. All rows of Table~\ref{tab:cal_offsets} are statistically independent, and we find offset calculations are in agreement to within $2.5\sigma$ between our work and that of both \citetalias{Brout_2022} and \citetalias{Popovic2025}. We note that currently DEBASS does not utilize the $z$-band in its light-curve fits \citep{nora}.

The offsets found here are an important crosscheck, as our measured offsets are largely statistically independent of \citetalias{Brout_2022} because we calibrate across the full DECam footprint as DEBASS is an all-sky program, whereas \citet{Brout_2022} only used tertiary stars in the dedicated DES SN fields. However, because we do not retrain the SALT3 model, we apply the \citetalias{Brout_2022} DES offsets as our nominal calibration. Additionally, because the observed differences in Table~\ref{tab:cal_offsets} are larger than the 7~mmag per band systematic uncertainty on the corrections reported in \citetalias{Brout_2022}, we apply the differences to \citetalias{Popovic2025} as a more conservative systematic uncertainty. We also propagate the per‑band calibration uncertainties following the uncertainty floor set in \citetalias{Brout_2022}, since our reduction and tertiary‑star cross‑calibration procedure follows the same methodology.

\section{Simulations}\label{sec:Sims}

Simulations are used to identify expected observational biases in our SN sample and correct for them. The structure of this section is as follows. In Section~\ref{sec:lcfit}, we summarize the SALT3 light-curve fitting that provides the input parameter distributions. We provide an overview of the simulations in Section~\ref{sec:simover}, address the selection criteria in Section~\ref{sec:selc}, and compare the simulated and observed DEBASS samples in Section~\ref{sec:simvdat}. Finally, Section~\ref{sec:bct} presents the resulting bias-correction analysis.

\subsection{Light Curve Fitting}\label{sec:lcfit}
To derive distances for our simulated SNe~Ia, we fit their multi-band light curves using the SALT3 model \citep{Kenworthy21} as implemented in the SuperNova ANAlysis software package, \git{rickkessler}{SNANA} \citep{snana}. In this analysis, we use the custom-trained \texttt{SALT3.DES5YR} model from \citet{vincenzi2024}, which follows the methodology of \citet{taylor2023} and is trained on the sample from \citet{Kenworthy21} using calibration values from \citet{Brout_2022}.

The SALT3 model returns five SN-dependent parameters: redshift ($z$), time of peak brightness ($t_{\rm peak}$), amplitude ($x_0$ or $m_x = -2.5 \log_{10}(x_0)$), stretch ($x_1$), and color ($c$). In our fitting, we fix redshift to the known host galaxy spectroscopic value as in \cite{nora}, and determine best-fit values and uncertainties for the remaining parameters via $\chi^2$ minimization. These parameters are converted into distance moduli using a modified version of the Tripp estimator \citep{Tripp98} following \citet{vincenzi2024}:
\begin{equation} \label{trip}
\mu = m_x + \alpha x_1 - \beta c - M - \Delta \mu_{\rm bias} + \Delta \mu_{\rm host}.
\end{equation}

\noindent Here, $\mu$ is the distance modulus; $\alpha$ and $\beta$ are global nuisance parameters describing the correlations between luminosity and light-curve shape ($x_1$) and color ($c$), respectively; and $M$ is the absolute magnitude of a fiducial SN~Ia with $x_1 = 0$ and $c = 0$. The term $\Delta \mu_{\rm host}$ is a correction for possible correlations between SN~Ia luminosity and host-galaxy properties, applied as a step function:
\begin{equation}
    \Delta\mu_{host} = \begin{cases}
+\gamma/2, & \text{if } M_{\star} < 10^{10}M_{\odot} \\
-\gamma/2, & \text{if } \text{otherwise}
\end{cases}
\end{equation}
where $\gamma$ is the size of the host mass step \citep{Lampeitl10, Sullivan10} and $M_{\star}$ is the stellar mass of the SN host galaxy. The distances and nuisance parameters ($\alpha$,$\beta$,$\gamma$) differ from those found in \cite{nora} because here we also implement the bias corrections ($\Delta \mu_{\rm bias}$) to account for biases due to selection effects and redshift-dependent detection efficiencies as determined by our simulations.

\subsection{Overview of Simulations}\label{sec:simover}
To evaluate the performance of our program in detecting and characterizing SNe~Ia, we run realistic simulations using \git{rickkessler}{SNANA} and the \git{dessn}{PIPPIN} pipeline architecture \citep{Hinton2020}. To model intrinsic scatter, we use the dust-based model from \citet{BroutScolnic21} (hereafter \citetalias{BroutScolnic21}). In addition to modeling SN~Ia parameters independently, we incorporate host-dependent parent populations by assigning SN~Ia properties based on the properties of their host galaxies following \cite{wiseman_2022}. This approach allows us to account for observed correlations between supernova parameters and host galaxy properties.

\texttt{SNANA} simulations take in three main inputs: (1) a source model for producing SNe with realistic astrophysical and cosmological properties, (2) a noise model that includes observational uncertainties based on program-specific conditions, and (3) a trigger model describing the probability of detection and spectroscopic confirmation based on program cadence and signal-to-noise. The combination of these components allows us to forward-model the underlying SN population and accurately estimate the selection bias in distance moduli, which is corrected using the $\Delta \mu_{\rm bias}$ term in our cosmological analysis.

Using information from the DEBASS light curves, we create simulations that match our observational conditions, including information such as the observation dates, zeropoints, point spread function, and sky background. We realistically model the DEBASS program cadence, convert supernova magnitudes into expected CCD counts, and calculate photometric uncertainties. 

A synthetic SN~Ia population is then generated using the SALT3 model parameters $z$, $x_1$, and $c$. Each simulated SN is matched to a host galaxy drawn from a galaxy catalog; we use the same galaxy catalog as DES-SN5YR, built with data \citep{qu2024}, which contains all of the galaxies detected using deep coadds with limiting magnitudes of $\sim 27$ in $g$-band \citep{Wiseman_2020}.

\subsection{Modeling the Selection Function}\label{sec:selc}
To generate DEBASS simulations, we strive to replicate the selection effects and follow-up efficiency of the DEBASS program, which we derive from the data. 

\begin{figure}
    \includegraphics[width=0.99\linewidth]{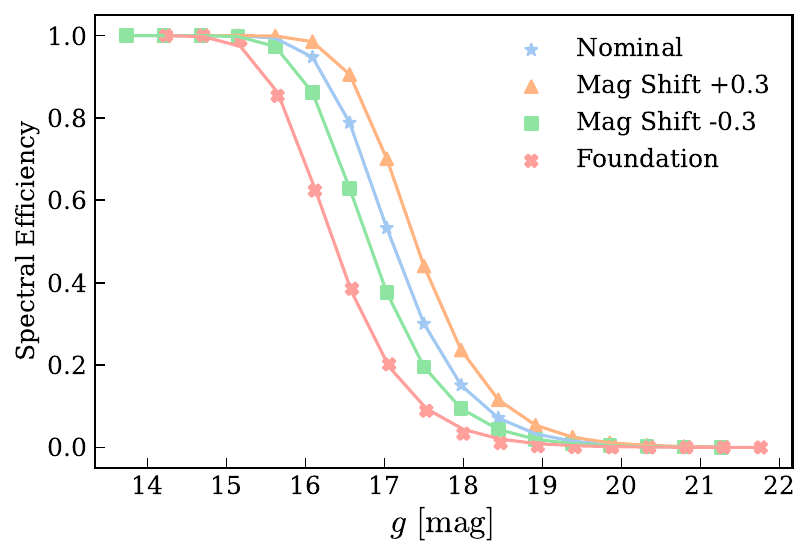}
    \caption{The spectroscopic selection efficiency as a function of peak $g$-band magnitude for DEBASS and Foundation.}
    \label{fig:eff}
\end{figure}

To model the selection efficiency, we compare DEBASS simulations with observed peak g-band magnitudes, finding that the relationship is well described by a sigmoid model. We then apply magnitude shifts to account for DEBASS-specific selection systematics and generate a series of modified selection functions by shifting this sigmoid in magnitude space across ten evenly spaced values ranging from $-0.3$ to $+0.3$~mag. For each shifted selection function, we run simulations and compare the resulting redshift distribution to that of the data using a $\chi^2$ fit. In Figure~\ref{fig:eff}, we show the selection function that yields the lowest $\chi^2$ value (blue stars) and is adopted as the best-fit model for our DEBASS simulations (see Figure~\ref{fig:6_hist}). The selection functions for Foundation, $-0.3$- and $+0.3$-magnitude shifts are shown as well; we adopt the $-0.3$~mag shift as a conservative estimate of the selection function systematic in Section~\ref{sec:hubres}, as it corresponds to approximately a $1\sigma$ deviation from the best-fit minimum in the $\chi^2$ distribution.

\subsection{Comparison of Simulations \& Data }\label{sec:simvdat}
\begin{figure}[!htb]
    \centering
    \includegraphics[width=0.9\linewidth]{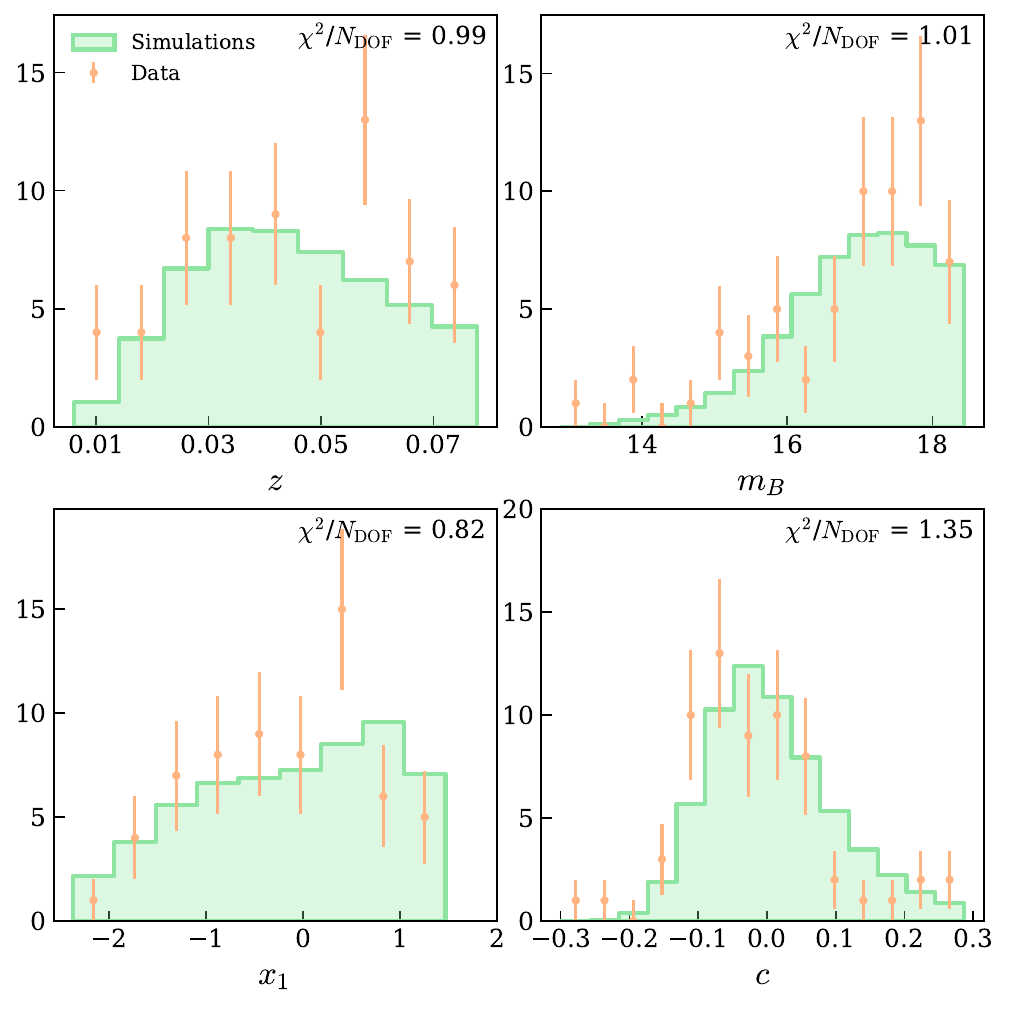}
    \caption{Comparison of data (orange dots) and simulations (green histogram) for distributions in the DEBASS sample, where the simulation is scaled to have the same number of events as the data. Redshift and stretch distributions are shown in the left column, while apparent magnitude and color are in the right column.}
    
    \label{fig:6_hist}
\end{figure}

We apply selection criteria based on the SALT3 light-curve fit parameters, signal-to-noise ratio, and light-curve coverage, following the approach in section 3.1.1 of \citet{nora}. After these cuts, we compare the DEBASS data and simulations in Figure~\ref{fig:6_hist}, finding overall good agreement.

In contrast to the Foundation sample, the DEBASS data show a noticeable deficit of SN~Ia values outside the $-1 <~ x_1~<1$ range \citep{nora}. To better match the observed distribution, we modify the SN~Ia stretch distribution in our simulations by reducing the probability of high-$x_1$ SNe by one-third across all host mass bins, relative to the distribution used in \citet{vincenzi2024}. It is unclear whether this trend reflects an intrinsic property of the full DEBASS sample or is a result of the specific selection of our current subsample. As such, we treat this modification as a source of systematic uncertainty in Section~\ref{sec:systondist}.

The color distribution of the DEBASS sample also differs from that of Foundation, with a tendency toward bluer values (a mean color of -0.02 for DEBASS compared to -0.01 for Foundation; \citealp{nora}). This may reflect a population of less-reddened, intrinsically bluer SNe~Ia or may be the result of selection biases favoring bright, blue transients in the discovery surveys that feed into the DEBASS program. However, unlike $x_1$, we do not introduce any modifications to the $c$ distribution in our simulations and do not currently treat it as a source of systematic uncertainty. We leave this effort for future work if this color discrepancy persists in the full 500 SN DEBASS sample. A more rigorous treatment of color would require modeling both intrinsic color and dust using approaches like Dust2Dust \citep{popovic2022}, which jointly fit the underlying color and dust distributions (including their host-mass dependence) through forward-modeling and reweighted selection efficiencies. 

\begin{table*}
    \centering
    {\fontsize{7pt}{10pt}\selectfont
    \setlength{\tabcolsep}{4pt}
    \caption{BBC-fitted nuisance parameters and the median difference between parameters (Med.~$\Delta p = \mathrm{median}(p_{\mathrm{Nominal}})-\mathrm{median}(p_{\mathrm{syst}})$) where $p$ is a given parameter) for a SN. All columns are in magnitudes except for $\alpha$, $\beta$, and $\gamma$, which are unitless.}
    \begin{tabularx}{\linewidth}{l@{\extracolsep{\fill}}ccccccccc}
    & $\sigma_{int}$ & $\alpha$ & $\beta$ & $\gamma$ & RMAD & Med.~$\Delta m_B$ & Med.~$\Delta \alpha x_1$ & Med.~$\Delta \beta c$ & Med.~$\Delta \mu$ \\\hline\hline
    
    Nominal &  $0.055$  &  $0.163 \pm  0.020$ & $ 3.375 \pm  0.217 $ & $0.064 \pm  0.038$ & $0.082$ & - & - & - & - \\\hline
    Acevedo ZP Offset & $0.059$  &   $0.164 \pm  0.020$ & $3.300 \pm  0.216$ & $0.065 \pm  0.036$ & $0.083$ & -$0.006 $ & $0.000 $ & $0.003 $ & -$0.009 $  \\
    Dovekie ZP Offset & $0.049$ & $0.165 \pm 0.020$ & $3.353 \pm 0.209$ & $0.065 \pm 0.037$ & $0.081$ & -$0.002 $ & $0.000 $ & -$0.026 $ & $0.024 $ \\
    Internal Calib.~ZP Offset & $0.054$ & $0.164 \pm 0.020$ & $3.366 \pm 0.205$ &  $0.064 \pm  0.037$ & $0.082$ & -$0.002 $ & $0.000 $ & -$0.006 $ & $0.004 $ \\
    Selection Function Shift~~ & $0.055$  &   $0.163 \pm  0.020$ & $3.385 \pm  0.215$ &  $0.065 \pm  0.037$ & $0.082$ & $0.000 $ & $0.000 $ & $0.000 $ & $0.000 $ \\
    Stretch Dist.~Change & $0.055$  & $0.163 \pm 0.020$ &  $3.369 \pm 0.217$ &  $0.063 \pm 0.038$ & $0.082$ & $0.000 $ & $0.000 $ & $0.000 $ & $0.000 $ \\
    $g$ Mean Filt.~Wav.~Shift & $0.052$ &  $0.163 \pm 0.021$ &  $3.362 \pm 0.207$ &  $0.067\pm 0.038$ & $0.082$ & -$0.001 $ & $0.001 $ & -$0.003 $ & $0.003 $ \\
    $r$ Mean Filt.~Wav.~Shift & $0.054$  & $0.163 \pm 0.020$ &  $3.369 \pm 0.217$ & $0.063 \pm 0.037$ & $0.080$ & -$0.000 $ & $0.000 $ & -$0.002 $ & $0.002 $ \\
    $i$ Mean Filt.~Wav.~Shift & $0.054$  & $0.163 \pm 0.020$ & $3.373 \pm 0.217$ & $0.064 \pm 0.038$ & $0.082$ & -$0.000 $ & $0.000 $ & -$0.003 $ & $0.003 $ \\\hline
        \label{tab:med_syst_shift}
    \end{tabularx}}
\end{table*}

\subsection{Bias Corrections}\label{sec:bct}
Our analysis uses bias corrections derived from large-scale SN~Ia simulations following the BEAMS with Bias Corrections (BBC) framework \citep{Kessler_2017, 2019kessler}. Here we refer to the ``BBC4D'' framework, as presented in \citet{Popovic_2021}. This method uses large-scale SN~Ia simulations that model the specific characteristics and selection effects of the programs included in the analysis. The BBC4D framework accounts for potential biases as a function of four key observables: $z$, $x_1$, $c$, and $\log M_\star$ (host-galaxy mass). The bias correction is defined as $\Delta\mu_{\text{bias}} = \mu - \mu^{\text{true}}$ where $\mu$ is defined in Equation~\ref{trip} and $\mu^{\text{true}}$ is the true cosmological distance modulus.

When using the dust-based forward-modeling approach of \citetalias{BroutScolnic21}, the method does not assume a single value of $\beta$. Instead, it simultaneously draws from distributions of intrinsic color-luminosity relations ($\beta_{\text{int}}$) and host-galaxy dust extinction laws. For the BBC4D bias correction, however, a single effective value of $\beta_{\text{true}}$ must be chosen. Given our small DR0.5 sample size for constraining $\beta_{\text{true}}$, and as \citet{Popovic_2021} showed that the precise choice of $\beta_{\text{true}}$ has a negligible impact on the final cosmological parameters, we adopt an effective value of $\beta_{\text{true}} = 2.87$ as in \citet{vincenzi2024}.

Supernova distances (and their uncertainties) that are corrected for selection biases and contamination are outputted from the BBC4D method. Distances are computed for both the nominal analysis and a range of alternative analysis configurations used to assess systematic uncertainties.

\section{Results}\label{sec:results}
In this section, we present the key results of our analysis, focusing on the impact of DEBASS-specific calibration and selection effects on distance measurements. We first quantify how the systematic distance offsets discussed in the previous sections propagate into the Hubble diagram. We then examine the Hubble residuals of DEBASS relative to Foundation. Finally, we assess the intrinsic Hubble scatter of the DEBASS sample and evaluate whether it is consistent with expectations from simulations.

\subsection{Calculating Systematic Effects on Distances}\label{sec:systondist}
To evaluate the robustness of our distance estimates, we quantify how various systematic effects impact our inferred distance moduli. We define our nominal sample as the DEBASS simulation that incorporates the Fragilistic zeropoint offsets, and the best-fit selection function as described in Section~\ref{sec:selc}. We then vary individual assumptions in turn to assess their impact on distances, holding all other components fixed. 

In Table~\ref{tab:med_syst_shift}, we summarize these systematic tests, with the following naming convention: (1) Acevedo ZP Offset: Applies the difference between the values in the first two rows of Table~\ref{tab:cal_offsets} (0.006, 0.008, 0.006, 0.018 in $griz$ respectively); (2) Dovekie ZP Offset: Applies external calibration offsets based on the \citetalias{Popovic2025} solution mentioned in Section~\ref{sec:extcalib} and taken to be a magnitude offset of 0.0015, 0.0014, 0.004, and -0.006 in $griz$; (3) Internal Calib.~ZP Offset: Applies the offset between our nightly calibration and DES (as listed in Section~\ref{sec:intcalib}); (4) Selection Function Shift: move the nominal selection function by -0.3~mag to test the sensitivity of bias corrections (Section~\ref{sec:selc}); (5) Stretch Dist.~Change: Adjusts the intrinsic distribution of SN~Ia stretch in the simulation to better match the observed distribution in our data (Section~\ref{sec:simvdat}); (6) Mean Filt.~Wav.~Shift: Alters the effective wavelength of each filter using 1$\sigma$ DES systematic uncertainties from table 2 in \citetalias{Popovic2025} (5, 5, 10 for $gri$).

We compute distances using the SALT3 light-curve parameters and use BBC4D output for the nuisance parameters ($\alpha$,$\beta$,$\gamma$) and distance moduli (see Section~\ref{sec:simover}). For each configuration, we calculate the median $\Delta\mu$, defined as the median difference in distance modulus (using Equation~\ref{trip}) compared to the nominal sample \textemdash most systematic variations shift distances by less than 1 millimagnitude. We also calculate the median difference in each term that goes into the distance modulus ($m_b$, $\alpha x_1$, $\beta c$), using a constant $\alpha=0.16$ and $\beta = 3.3$ and report the intrinsic scatter ($\sigma_{\mathrm{int}}$) in the Hubble Diagram, as well as the robust median absolute deviation of the Hubble residuals ($\mathrm{RMAD}$) defined in \cite{Hoaglin00} as:
\begin{multline}\label{eq:rmad}
    \mathrm{RMAD} = 1.48 \times \mathrm{median}(|(\mu - \mu_{\mathrm{model}})- \\\mathrm{median}(\mu - \mu_{\mathrm{model}})|).
\end{multline} 
\noindent We find $\sigma_{\mathrm{int}} \sim 0.06$~mag, lower than that reported by \citetalias{BroutScolnic21} and consistent with the higher signal-to-noise ratio of the DEBASS dataset. Finally, we find that the nuisance parameters $\alpha$, $\beta$, and $\gamma$ remain stable across all tested systematics where $\alpha \sim 0.16$, $\beta \sim 3.4$, and $\gamma \sim 0.06$~mag. The $\mathrm{RMAD}$ also remains constant across all systematics at $\sim 0.08$, which is lower than the $0.1$ value from \citet{nora} due to the use of bias corrections.

The largest term comes from the Dovekie calibration offset, which shifts the median $\Delta \mu$ by $+0.024 \pm 0.001$~mag, which is dominated by the color calibration ($\Delta \beta c$ of -0.026), reflecting a full recalibration using updated zeropoint offsets \citep{Popovic2025}. 
The difference between \citetalias{Brout_2022} and \citetalias{Popovic2025} is much larger than the difference found for the recalibration performed in this work ($-0.009$~mag for the Acevedo ZP offsets and $+0.004$~mag for the internal calibration offsets). 
We note that Table~\ref{tab:med_syst_shift} considers these calibration changes in isolation for DEBASS, while \citetalias{Popovic2025} includes recalibration for other programs that must also be considered (See Section~\ref{sec:hubres}).

We also include the selection function shift and stretch distribution change because they probe our ability to model the selection function and host-galaxy-dependent parent populations accurately. These are found to be much smaller in value compared to calibration. Filter wavelength shifts also contribute at the $0.002$–$0.003$~mag level, and are found to be subdominant. The distance-offset budget is therefore dominated by calibration.

\subsection{Hubble Residuals}\label{sec:hubres}
\begin{figure}
    \includegraphics[width=0.99\linewidth]{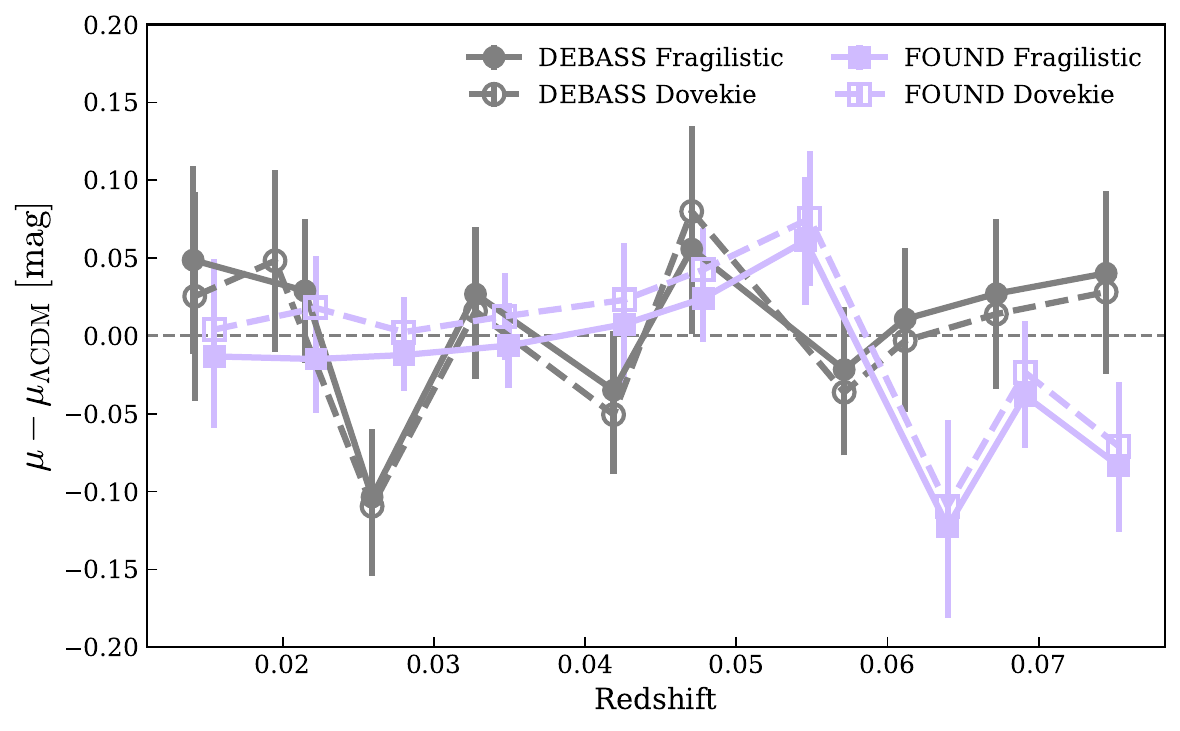}
    \caption{Redshift vs.~the median of the Hubble residuals ($\mu - \mu_{\Lambda \mathrm{CDM}}$) for the two low-$z$ programs using either \citetalias{Brout_2022} (filled points) or \citetalias{Popovic2025} (unfilled points).}
    \label{fig:mu_surv}
\end{figure}

In Table~\ref{tab:medres} and Figure~\ref{fig:mu_surv}, we compare the median Hubble residuals of the DEBASS and Foundation samples. We ensure that DEBASS and Foundation are reanalyzed consistently and examine the Hubble residual differences between DEBASS and Foundation for both the \citetalias{Brout_2022} calibration and for \citetalias{Popovic2025}. These residuals in Table~\ref{tab:medres} represent the median difference between the observed and $\Lambda$CDM model-predicted distance moduli between the two surveys, computed after applying the BBC4D bias correction method (see Section~\ref{sec:bct}). For consistency, we reprocessed the Foundation dataset using SALT3 and the \citetalias{BroutScolnic21} intrinsic scatter model to match the DEBASS simulation framework. As shown in Table~\ref{tab:medres}, the overall median offsets between DEBASS and Foundation are small, and changing the calibration to \citetalias{Popovic2025} only changes the direction of the shift.

\begin{table}[]
    \caption{Median difference of the Hubble residuals ($\mu - \mu_{\Lambda \mathrm{CDM}}$~[mag]) between DEBASS and Foundation processed with \citetalias{Brout_2022} calibration and \citetalias{Popovic2025} calibration.}
    \centering
    \begin{tabularx}{\columnwidth}{c@{\extracolsep{\fill}}c}
        ~~~Fragilistic Calibration & Dovekie Calibration~~ \\ \hline\hline
        $0.016 \pm 0.019$ & $-0.015 \pm 0.018$ \\ \hline
    \end{tabularx}
    \label{tab:medres}
\end{table}

Figure~\ref{fig:mu_surv} presents the binned Hubble residuals as a function of redshift for DEBASS and Foundation. Each point shows the median distance residual ($\mu - \mu_{\Lambda \mathrm{CDM}}$) in a redshift bin, with error bars representing the inverse-variance weighted uncertainty of the bin. The residual trends with redshift are flat within uncertainties, with scatter at the $\sim$0.02--0.05~mag level across bins. No systematic evolution with redshift is apparent, reinforcing that the two samples are in good agreement.

As shown in Table~\ref{tab:medres}, applying the \citetalias{Popovic2025} calibration shifts both DEBASS and Foundation by nearly the same amount, but in opposite directions relative to the \citet{Brout_2022} calibration used in DEBASS. This behavior demonstrates that the relative difference between the programs remains robust, regardless of the absolute calibration framework.

\subsection{Hubble Scatter}\label{sec:hubscat}
The achromatic component of Hubble diagram scatter ($\sigma_{\rm int}$) is determined by adjusting its value until the reduced $\chi^{2}$ relative to a best-fit cosmology equals 1. We note that the value for $\sigma_{\rm int}$ determined in this paper differs from that of \cite{nora} because here we have included the tuning of uncertainties by our bias-correction simulations following \cite{Brout22}, which accounts for increased uncertainties for the reddest SNe. This results in a $\sigma_{\rm int}$ of 0.055~mag as shown in Table~\ref{tab:med_syst_shift}. In contrast, the Hubble scatter is computed as the $\mathrm{RMAD}$ of the Hubble residuals (as shown in Equation~\ref{eq:rmad}). While $\sigma_{\rm int}$ represents the smallest possible scatter after accounting for measurement and host-dust modeling uncertainties, the Hubble scatter is larger as it describes the full dispersion of Hubble residuals for the dataset.

\begin{figure}
    \includegraphics[width=0.99\linewidth]{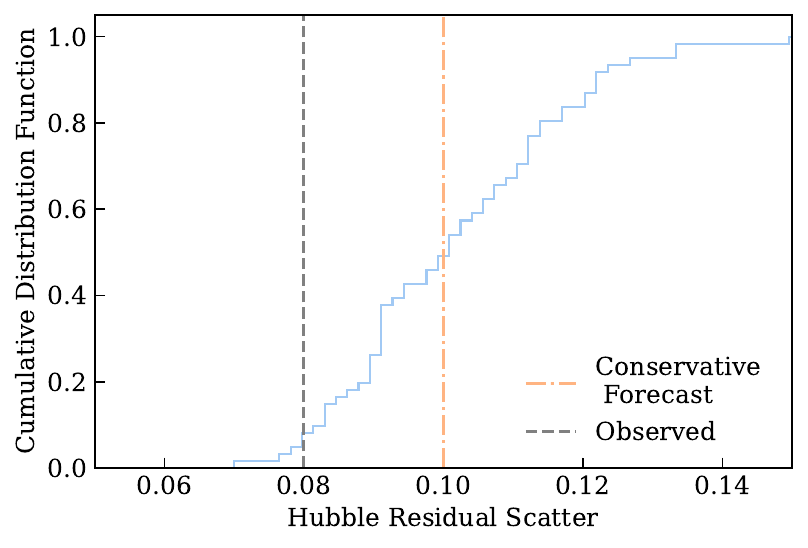}
    \caption{Histogram of Hubble residual scatter ($\mu - \mu_{\Lambda \mathrm{CDM}}$ with bias corrections where $\mu_{\Lambda \mathrm{CDM}}$ is the expected value at a given redshift in a $\Lambda$CDM cosmology for simulations with $62$ SNe~Ia). Dashed lines show the Hubble scatter found in this paper (grey) as well as the conservative estimate used in our forecasts (orange).}
    \label{fig:schunk}
\end{figure}

To evaluate the reliability of our low observed $\mathrm{RMAD}$ of $\sim$0.08~mag seen in Table~\ref{tab:med_syst_shift} ($\sim$0.1~mag found in \citet{nora} without bias corrections), we analyze realistic, bias-corrected simulations of DEBASS. To interpret this value, we divide the simulated sample into subsets that match the DEBASS sample size of 62 cosmologically ready SNe from \citet{nora} and examine the distribution of their resulting Hubble scatters. Figure~\ref{fig:schunk} shows a histogram of the simulated Hubble residual scatter, and we find that our conservative residual scatter occurs in approximately 10\% of these subsets, indicating that the observed scatter is statistically consistent with the expectations of our simulation framework. Nonetheless, to produce the conservative forecasts of Section~\ref{sec:forecast}, we will use $0.1$, which corresponds to the scatter from \citet{nora}. 

\section{Forecasts}\label{sec:forecast}
We assess how well the full DEBASS supernova dataset can constrain two key cosmological parameters: the dark energy equation-of-state parameter, $w$, and the product of the growth rate of cosmic structure and the amplitude of matter density fluctuations, denoted as $f\sigma_8$. Though the current DEBASS sample is over 500 SNe and growing, we conservatively conduct our forecasts assuming 400 SNe that pass cosmology cuts.

\subsection{Cosmic Expansion History Constraints}\label{sec:ffor:de}
In the standard cosmological model, dark energy is often described by its equation-of-state parameter $w$, which relates pressure to energy density. An evolving dark energy is parametrized by $w(a)=w_0+w_a(1-a)$, where $w_0$ represents the present-day value and $w_a$ captures its possible evolution with scale factor $a$. Alongside these, the matter density parameter ($\Omega_M$) quantifies the present-day fraction of the Universe’s energy density in matter. 

To evaluate the impact of the low-redshift SN~Ia sample on cosmological constraints when combined with the DES-SN5YR high-$z$ sample \citep{vincenzi2024}, we perform a Fisher matrix analysis to estimate the uncertainties on $w_0$, $w_a$, and $\Omega_M$. We compare the performance of DEBASS against the historical low-$z$ sample used in the previous analyses of the DES-SN5YR program.

We follow the Fisher formalism as described in \cite{Tegmark_1997}. We define our data vector $\boldsymbol{V}$ as:
\begin{equation}
     \boldsymbol{V} = \mu_\text{model}(\boldsymbol{z}; \boldsymbol{\Omega}_\text{model}) + M_0,
\end{equation}
where $\mu_\text{model}$ is the distance modulus in the considered cosmological model evaluated at the redshifts $\boldsymbol{z}$ of the sample, $\boldsymbol{\Omega}_\text{model}$ are the free parameters of the cosmological model and $M_0$ is an offset that represents the absolute magnitude of SNe~Ia and which is included during the Fisher matrix computation to properly capture degeneracies.

The Fisher matrix for the set of parameters $\boldsymbol{\Theta} = \left\{\boldsymbol{\Omega}_\text{model}, M_0\right\}$ is defined by:
\begin{equation}
F_{ij} = \boldsymbol{V}_{,i}^\text{T} C^{-1}  \boldsymbol{V}_{,j},
\end{equation}
where $\boldsymbol{V}_{,i} = \frac{\partial}{\partial\Theta_i}\boldsymbol{V}$and $C$ is the covariance matrix.
In the following, the derivatives are estimated at the best-fit parameters of the DES-SN5YR analysis \citep{descollaboration2024darkenergysurveycosmology}.

We consider two simulated samples for our dark energy forecasts, the first is the one used in DES-SN5YR analysis and consists of SNe~Ia observed by the DES survey combined with multiple low-redshift surveys (193 low-$z$ SNe Ia); we refer to this sample as DES+Historic. In the second sample, we replace the combination of low-redshift surveys with the DEBASS sample, which consists of $O(400)$ SNe~Ia; we refer to this sample as DES+DEBASS. Beyond roughly doubling the number of low-$z$ SNe, DEBASS provides a significantly lower Hubble residual scatter ($\sim0.10$~mag) compared to the Historic sample ($\sim0.12$~mag), leading to tighter distance constraints. The covariance matrix of the DES+Historic dataset is a diagonal matrix using the statistical errors estimated in \cite{descollaboration2024darkenergysurveycosmology}. For the DES+DEBASS sample, the covariance is also a diagonal matrix where the errors of DES SNe~Ia are the same as for the DES+Historic sample, and the errors for the DEBASS sample are extrapolated from the \citetalias{BroutScolnic21} model. This is a consistent comparison with DES+Historic because in DES+Historic, the \citetalias{BroutScolnic21} model was also used to rescale the reported uncertainties. Likewise, the errors for DES+DEBASS are obtained using a large simulation to compute the dependence of the scatter on the color $c$. 

\begin{figure}
    \centering
    \includegraphics[width=\linewidth]{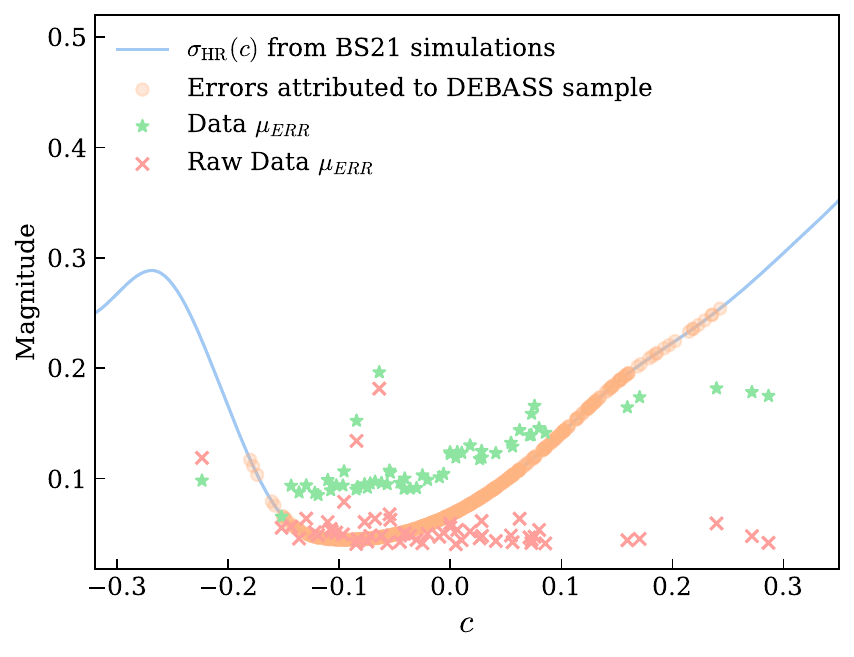}
    \caption{Color dependence of the Hubble diagram scatter predicted from the \citetalias{BroutScolnic21} model. The blue line is obtained from a large simulation. Orange point represents the error values attributed to DEBASS sample and used in our Fisher forecast. Green stars represent the bias-corrected errors in $\mu$ for the DEBASS sample, and the red crosses are the uncorrected errors.}
    \label{fig:sigofc}
\end{figure}

When then assign color for the DEBASS sample according to the distribution defined in \cite{popovic2022}. Corresponding errors are obtained by interpolation of the $c$ parameters on the relation found from the simulation. Finally, we rescale all the errors with a constant such that the mean error of our sample is equal to the current scatter value of DEBASS, following the simulated curve of Figure~\ref{fig:sigofc}. In this figure, the blue curve shows the simulated intrinsic scatter $\sigma_{\rm HR}(c)$, while the raw photometric uncertainties of our sample are shown as red crosses. The fact that the photometric errors lie at or below the intrinsic scatter curve indicates that the uncertainties arise primarily from the natural variability of SNe, with minimal contribution from calibration or model systematics.

\begin{figure}
    \centering
    \includegraphics[width=\linewidth]{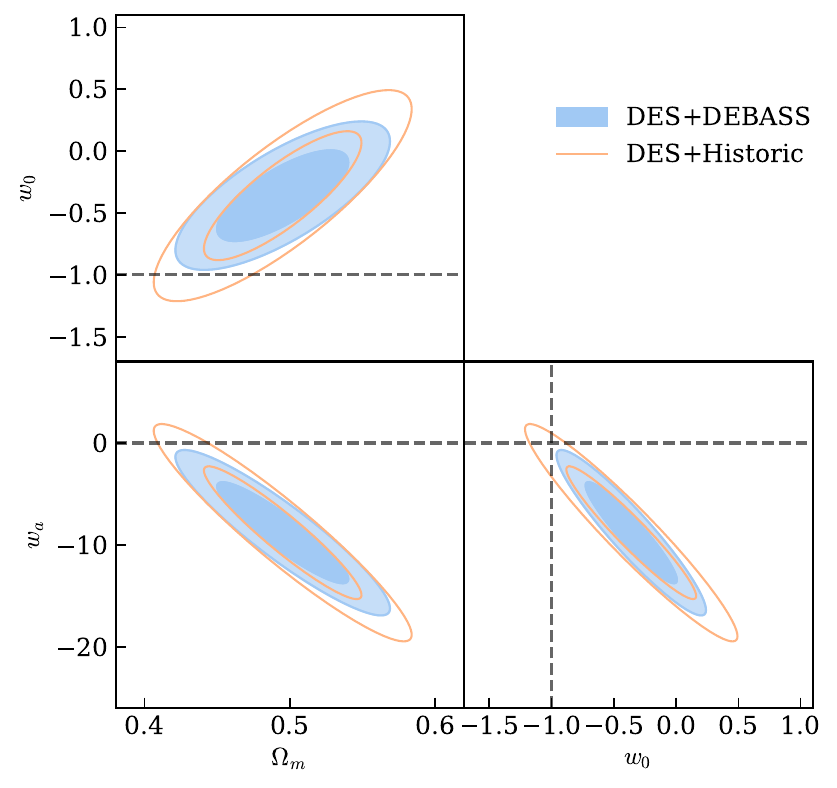}
    \caption{Fisher forecast $\Omega_m-w$ contours for the $w_0w_a$CDM model. The two contours represent the 68.3\% and 95.4\% confidence levels. Blue and orange contours are the Fisher forecast for the DES+DEBASS and DES+Historic samples, respectively. The dashed lines represent the values for $\Lambda$CDM.}
    \label{fig:Omw}
\end{figure}

\begin{table*}
    \caption{Results for the Fisher forecast of the DES+DEBASS sample for the different cosmological models. Results between parentheses are the Fisher forecast from the DES+Historic sample.}    \label{tab:fmod}

    \centering
    \renewcommand{\arraystretch}{1.2}
    \begin{tabularx}{\textwidth}{c@{\extracolsep{\fill}}|cccc}
    Model              & $\sigma_{\Omega_M}$ & $\sigma_{\Omega_k}$ & $\sigma_{w_0}$ & $\sigma_{w_a}$  \\ \hline\hline
    Flat-$\Lambda$CDM  & 0.008 (0.011)       & -                   & -              & -               \\
    $\Lambda$CDM       & 0.049 (0.053)       & 0.117~(0.134)       & -              & -               \\
    Flat-$w$CDM        & 0.070 (0.077)       & -                   & 0.119 (0.139) & -               \\
    Flat-$w_0w_a$CDM~~~~~~   & 0.030 (0.036)       & -                   & 0.242~(0.344)  & 3.27 (4.30)     \\ \hline
    \end{tabularx}
\end{table*}

We consider the impact of the DEBASS sample on a range of cosmological models, including Flat-$\Lambda$CDM, $\Lambda$CDM, Flat-$w$CDM, and Flat-$w_0w_a$CDM. 
Table~\ref{tab:fmod} shows that, across all models, we find that replacing the Historic low-$z$ sample with DEBASS leads to consistent improvements in cosmological parameter constraints. In the Flat-$w_0w_a$CDM model, the uncertainty on $w_0$ is reduced by approximately 30\% while the uncertainty on $w_a$ is reduced by about 24\%. This is shown in Figure~\ref{fig:Omw}, where we draw the marginalized posterior contours for the 68.3\% and 95.4\% confidence levels. The contours for DES+Historic and DES+DEBASS are plotted to directly compare the constraining power of the two different low-$z$ samples when paired with the DES-SN5YR high-$z$ dataset.
These improvements are driven by DEBASS’s precise calibration, higher signal-to-noise light curves, and a well-characterized selection function—all of which contribute to lower $\sigma_\text{int}$ and lower uncertainties. 

\begin{figure}
    \centering
    \includegraphics[width=0.99\linewidth]{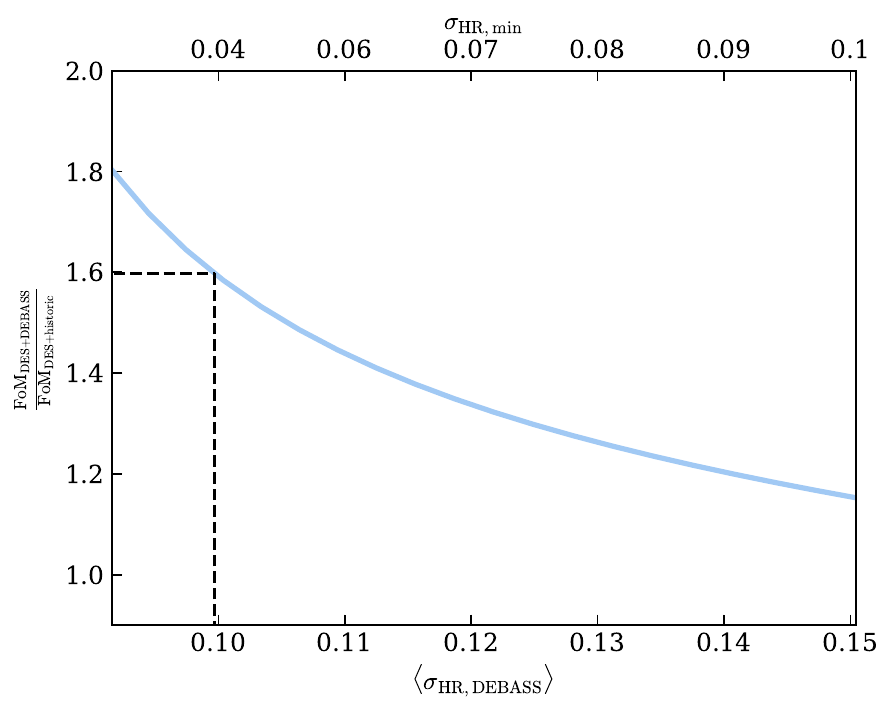}
    \caption{Ratio between the Fisher $w_0$-$w_a$ FoM of DES+DEBASS and DES+Historic samples as a function of the DEBASS scatter (bottom axis) and the minimum possible scatter (top axis). The dashed line indicates the current estimated scatter.}
    \label{fig:FoMevol}
\end{figure}

We also explore how the dark energy Figure of Merit (FoM), defined as $\mathrm{FoM} \propto 1/\left[\sqrt{\det \text{Cov}(w_0, w_a)}\right]$, evolves with assumptions about the intrinsic scatter value of the DEBASS sample \citep{Albrecht_2006}. In Figure~\ref{fig:FoMevol}, we show how the evolution of the ratio between the DES+Historic FoM and the DES+DEBASS FoM varies as a function of the minimum Hubble residual scatter, $\sigma_{HR, min}$(top axis), and the corresponding mean sample scatter (bottom axis). As shown in Figure~\ref{fig:FoMevol}, the conservatively estimated DEBASS scatter (indicated by the dashed line) yields a $\sim 60\%$ improvement on the FoM. Even with a high scatter of 0.13, the improvement on the FoM is $\sim 25\%$.

\subsection{Growth Rate of Structure}
The measurement of $f\sigma_8$ is a powerful tool to probe dark energy and gravity. Particularly, its precise constraint will enable tests of general relativity at large scales and help to discriminate between dynamic dark energy and modified gravity models \citep{houCosmologicalProbesStructure2023,huterer_growth_2023}.

To forecast the constraining power of the DEBASS sample, we follow the same Fisher formalism as used in the previous section and apply it to the maximum-likelihood method as developed in \citet{carreres_growth-rate_2023} and implemented in the \git{corentinravoux}{flip} Python library \citep{ravoux_generalized_2025}. 
We compute the velocity power spectrum $P_{vv}(k)$ with the  \gitalias{lesgourg}{class\_public}{CLASS} Boltzmann solver, we used the Planck18 $\Lambda$CDM results \citep{Planck18} as the input fiducial cosmology. The redshift space distortions are modeled by multiplying the velocity power spectrum with the damping factor $D_u(k,\sigma_u)=\text{sinc}(k\sigma_u)$ as described in \cite{Koda2014}. We fixed the value of the damping scale, $\sigma_u$, to $21~\mathrm{Mpc}~h^{-1}$ as discussed in \cite{carreresTypeIaSupernova2025}. A non-linear velocity dispersion parameter  $\sigma_v=300~\mathrm{km\,s^{-1}}$ is added on the diagonal of the covariance matrix to represent unresolved random motions.

\begin{figure}
    \centering
    \includegraphics[width=\linewidth]{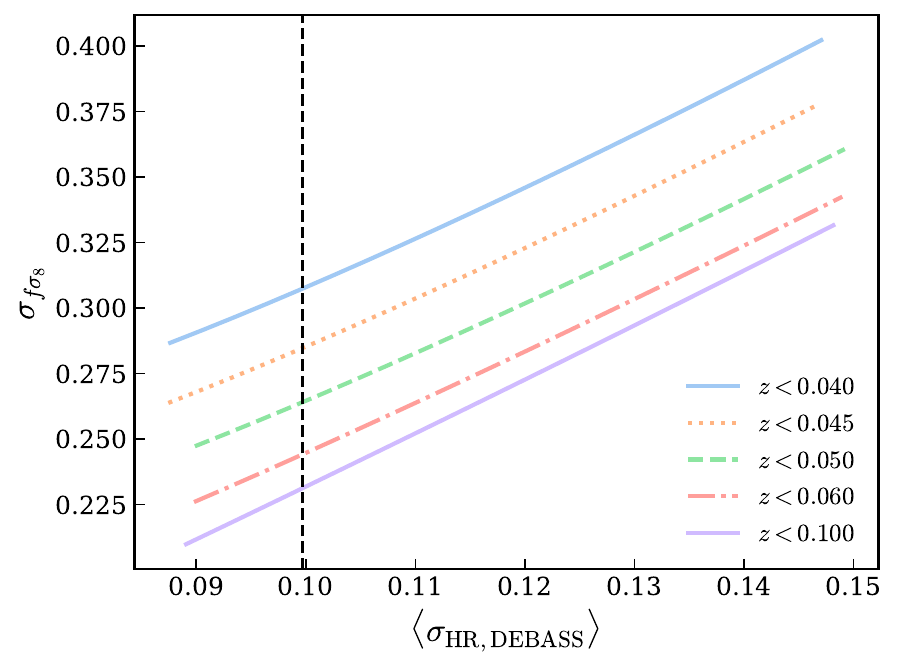}
    \caption{Fisher forecast for the growth rate of structure, $f\sigma_8$ as a function of the mean Hubble residual scatter. The forecast is done for several maximum redshift cuts. The vertical dashed line corresponds to the conservative estimate of the scatter of the DEBASS sample.}
    \label{fig:ff_fs8}
\end{figure}

The errors are modeled in the same way as in Section~\ref{sec:ffor:de}. Figure~\ref{fig:ff_fs8} shows the resulting forecast on $f\sigma_8$ as a function of the scatter of the DEBASS sample, and for different redshift cuts. The redshift cut will depend on the final selection function since Malmquist bias is known to bias the resulting $f\sigma_8$ \citep{carreres_growth-rate_2023}. However, the possibility to correct for Malmquist bias without affecting the $f\sigma_8$ measurement is currently being investigated and could enable the use of the whole DEBASS sample in our future analysis.
Figure~\ref{fig:ff_fs8} shows that we can expect a measurement of $f\sigma_8$ with a relative precision of $\sim 23 - 30\%$. Current constraints on $f\sigma_8$ at low redshift are slightly above 20\% precision. The DESI PV forecast from \cite{desipvf} for $0 \le z \le 0.05$ is $\sim 16-27\%$, which is derived using a power spectrum-based Fisher-matrix approach that yields systematically smaller predicted uncertainties than our field-level Fisher method (see, e.g., figure 10 of \citealt{ravoux_generalized_2025}). Looking ahead, the Vera C. Rubin Observatory’s Legacy Survey of Space and Time (LSST) is expected to achieve errors of 10–18\% on the growth-rate parameter after ten years of SNe~Ia observations \cite{rosselli2025forecastgrowthratemeasurementusing}. Our forecasts are thus somewhat less precise than current best measurements and optimistic DESI PV projections, but remain within a factor of $\lesssim 2$ of the precision expected from LSST at full survey depth.

\section{Conclusions}\label{sec:Conclusions}

In this paper, we build simulations and examine potential sources of systematic uncertainty affecting the DEBASS DR0.5 sample, which represents the first subset of a much larger low-redshift SN~Ia program ($>500$ SNe~Ia). Accurate treatment of these systematics is critical because low-$z$ samples serve as the anchor for cosmological distance measurements. We focus on two leading contributors: photometric calibration and the spectroscopic selection function. To address calibration, we implement multiple complementary strategies, including internal calibration using nightly observations and external cross-checks against DELVE and DES catalogs. These approaches ensure consistency across the DECam system and allow us to quantify calibration differences relative to other programs. Rather than assuming negligible offsets, we adopt a conservative approach that incorporates zeropoint adjustments from independent calibration efforts, as well as systematic variations in filter transmissions and color terms.

To model the program selection function, we use simulations to match the redshift and magnitude distributions of the DEBASS sample. We also identify a difference in the stretch distribution, with DEBASS containing fewer $x_1 > 1$ SNe~Ia than expected from Foundation. We address this by modifying the $x_1$ probability distribution as a function of host-galaxy mass, and we conservatively treat this as a systematic uncertainty in our simulation framework. Additionally, we note that the $c$ distribution in DEBASS appears bluer than in other low-$z$ samples, but it is unclear whether this trend reflects intrinsic properties or results from small-number statistics. A more sophisticated approach, such as Dust2Dust modeling \citep{popovic2022}, will be necessary in future work to accurately capture host-dependent color distributions and their potential impact on cosmological measurements.
From our bias-corrected Hubble diagram, we measure a SALT3 $\alpha$ parameter of $0.16 \pm 0.02$ and a $\beta$ parameter of $3.38 \pm 0.22$, consistent with previous low-$z$ SN~Ia analyses. We estimate an intrinsic scatter of $\sigma_{\rm int} = 0.055$~mag, which is lower than typical values reported in the literature. Our simulations reproduce this level of scatter, suggesting that the higher signal-to-noise ratio of DEBASS photometry makes such low scatter plausible. When comparing DEBASS to Foundation, we find excellent consistency, with a total systematic uncertainty of $0.024$~mag, dominated by photometric calibration. 

As a cross-check, we compare the DEBASS Hubble residuals to those from the Foundation sample \citep{Foley18}. Using consistent light-curve fitting and bias correction, we find a median Hubble residual offset of $0.016 \pm 0.019$~mag between DEBASS and Foundation when processed with the \citetalias{Brout_2022} calibration. When we switch to the \citetalias{Popovic2025} calibration, this offset becomes $-0.015 \pm 0.018$~mag. The offsets are of the same order but in different directions, which highlights the need for careful calibration treatment when combining datasets for cosmological inference.

By relying on a single, internally consistent low-$z$ dataset, DEBASS removes the need to model cross-program systematics inherent in the heterogeneous historic sample. In this analysis, we consider only statistical uncertainties to isolate the impact of sample homogeneity and measurement precision on cosmological constraints. Based on conservative simulation estimates, replacing existing historical low-redshift samples with the complete DEBASS dataset ($>$400 SNe~Ia) is expected to reduce statistical uncertainties on $w_0$ and $w_a$ by roughly 30\% and 24\%, respectively, boost the dark energy Figure of Merit by as much as 60\%, and deliver a $\sim$25\% precision measurement of $f\sigma_8$.

We have highlighted the key areas of calibration and selection systematics that impact our current analysis and established a framework for quantifying their influence on cosmological measurements. We are on track to deliver high-precision cosmological measurements from the DEBASS program. As the sample size grows and further improvements are made to calibration and modeling, DEBASS will provide an essential low-redshift anchor for next-generation SN~Ia cosmology.

\begin{acknowledgements}

\section*{Acknowledgements}
We thank the Templeton Foundation for directly supporting this research (N.S., D.B., and D.S.). D.S. is supported by Department of Energy grant DE-SC0010007, the David and Lucile Packard Foundation, and the Templeton Foundation. D.S. and M.A. are supported by the Alfred P. Sloan Foundation. This material is based upon work supported by the National Science Foundation Graduate Research Fellowship under Grant No. DGE 2139754. We thank the reader. We also thank Tamara Davis and Chris Lidman for their careful reading of this paper.

\section*{Author Contributions}
M.~A. performed the analysis and wrote the majority of the paper. N.~S. helped process the observed data and formulate the calibration. D.~B. guided the methodology and implementation of calibration, simulations, bias corrections, and systematics. PI of DEBASS. D.~S. advised and provided critical review and editing. B.~C. contributed to the forecasts and simulation aspects of the analysis and participated in the review and writing. B.~P. contributed guidance on photometric calibration and provided support for working with Dovekie. All authors provided editing and discussions around the paper.

\section*{Software}
astropy \citep{astropy:2013,astropy:2018,astropy:2022}, CLASS \citep{Diego_Blas_2011}, Dovekie \citep{Popovic2025}, flip \citep{ravoux_generalized_2025}, jax \citep{Campagne_2023}, matplotlib \citep{Hunter07}, numpy \citep{numpy11},
\texttt{pandas} \citep{the_pandas_development_team_pandas-devpandas_2024}, \texttt{PIPPIN} \citep{Hinton2020}, scipy \citep{scipy}, seaborn \citep{Waskom2021}, \texttt{SNANA} \citep{Kessler09}.

\end{acknowledgements}

\bibliographystyle{mn2e}
\bibliography{main}{}

\end{document}